\documentclass[12pt]{article} 
\usepackage[numbers]{natbib}
\usepackage{a4}
\usepackage{graphicx}% Include figure files
\usepackage{bm}
\usepackage{amsmath}
\usepackage{amssymb}
\usepackage{latexsym}
\usepackage{amssymb}
\usepackage{epsfig}
\usepackage{natbib}
\def \de{{\mathrm{d}}}

\def \pd{\partial}

\begin{document}
%%%%%%%%%%%%%%%%%%%%%%%%%%%%%%%%%%%%%%%%%%%%%%%%%%%%%%%%%%%%%%%%%%%%%
\title{{\bf 
A note on Lorentz-like transformations and superluminal motion
}}
\author{Congrui Jin~$^\text{a,}$\footnote{%%%Corresponding author. 
{\it E-mail address:} cj263@cornell.edu.}
\ 
and 
Markus Lazar~$^\text{b,}$\footnote{%%%%%Corresponding author. 
{\it E-mail address:} lazar@fkp.tu-darmstadt.de.}
%%%\newline
%%%\ Tel.:+49(0)6151/163686.%%%% Fax.: +49(0)6151/163681.
\\ \\
${}^\text{a}$ 
Field of Theoretical and Applied Mechanics, \\
Sibley School of Mechanical and Aerospace Engineering, \\
Cornell University, \\
Ithaca, NY 14853, USA\\
\\
${}^\text{b}$ 
        Heisenberg Research Group,\\
        Department of Physics,\\
        Darmstadt University of Technology,\\
        Hochschulstr. 6,\\      
        D-64289 Darmstadt, Germany
}

\date{\today}    
\maketitle
%%%%%%%%%%%%%%%%%%%%%%%%%%%%%%%%%%%%%%%%%%%%%%%%%%%%%%%%%%%%%%%%%%%%%%%%%%%%%

%%%\date{\today}
%%%\newpage

\begin{abstract}
In this extended note a critical discussion of an extension of the Lorentz transformations 
for velocities faster than the speed of light given recently by~\citet{HC} is provided. 
The presented approach reveals the connection between faster-than-light speeds 
and the issue of isotropy of space.
It is shown 
if the relative speed between the two inertial frames $v$ is greater than the speed 
of light, the condition of isotropy of space cannot be retained.  
It further specifies the respective transformations 
applying to $-\infty<v<-c$ and $c<v<+\infty$.
It is proved that 
such Lorentz-like transformations are improper transformations
since the Jacobian is negative.
As a consequence, the wave operator, the light-cone and 
the volume element are not invariant under such Lorentz-like transformations.
Also it is shown that such Lorentz-like transformations are not 
new and already known in the literature.
\\

\\
\end{abstract}
%%%\pacs{03.30}
% PACS: 03.30.+p,11.30.Cp,
% PACS, the Physics and Astronomy
 %%%\keywords{special relativity; superluminal; Lorentz transformation}

\newpage
\section{Introduction}
The early study of superluminal motion (motion faster than the speed of light) 
may be dated back to~\citet{Heaviside} and \citet{Sommerfeld} (see also~\citep{Schott}).
Recently, there has been a renewed interest in superluminal propagation in physics
(e.g.~\citep{Fayngold,Visser,CS,Vieira,HC}).
In this connection, 
\citet{HC} have introduced two possible ``new'' transformations to extend Lorentz transformations 
for inertial observers moving faster than the speed of light with respect to each other.
Unlike the usual Lorentz transformations, 
the one parameter family of transformations derived by~\citet{HC} 
for the regime $c<v<+\infty$ do not form a one parameter group of transformations, 
since neither the identity characterized by $v=0$ nor the inverse characterized by $-v$ 
are in the regime $c<v<+\infty$, and the extensions to negative $v$ need to be obtained 
from other possible approaches.

In this work, we show that generalized Lorentz transformations can be derived 
formally 
from the connection between faster-than-light speeds and the issue of isotropy of space.
On the other hand, we point out that the ``Hill-Cox transformations'' are not new 
since such transformations, sometimes called ``Goldoni transformations'', 
were earlier given in the 1970s and 1980s. 
In addition, we show that such generalized Lorentz transformations violate
the condition of isotropy of space. 
%%%and as a consequence the principle of relativity is violated.
For simplicity, most of the mathematical 
discussion will be restricted to the case of $(1+1)$-dimensions
(for $(3+1)$-dimensions, one has to add: $y'=y$, $z'=z$ 
if the relative velocity is parallel to the $x$-axis).
The modest goal of the present paper
is not to derive a theory of superluminal motion
but rather to discuss existing and recently published
generalized Lorentz transformations.

\section{Generalized Lorentz transformations}
It is assumed that we have two inertial frames, $S$ and $S'$, 
moving with relative velocity $v$ in the $x$-direction. 
Both inertial frames come with Cartesian coordinates: $(x, t)$ for $S$ and $(x', t')$ 
for $S'$. 
The most general possible relationship should be of the form: $x' = f(x, t)$, and $t' = g(x, t)$ 
for some functions $f$ and $g$. 
Since $S$ and $S'$ are both inertial frames, the map $(x, t) \to (x', t')$ 
must map straight lines to straight lines, that is, such maps are linear. 
$f$ and $g$ thus should take the form                
\begin{equation}
\label{single}
\left\{ \begin{array}{ccc}
x'\\
t' \\
\end{array} \right\}=\left[ \begin{array}{ccc}
a_1& a_2\\
a_3& a_4
\end{array} \right]\left\{ \begin{array}{ll}
x\\
t\\
\end{array} \right\}\,,
\end{equation}
with $a_i$, $i = 1$, $2$, $3$, $4$ each being a function of $v$, $a_i=a_i(v)$. 
Assume that $S'$ is traveling at velocity $v$ relative to $S$, 
which means that an observer sitting at $x' = 0$ of $S'$ moves along the trajectory $x = vt$ 
in $S$, that is, $x = vt$ must map to $x' = 0$. 
There is actually one further assumption that $x' = 0$ coincides with $x = 0$ at $t = 0$. 
Together with the requirement of linearity, this restricts the coefficients $a_1$  and $a_2$ 
to be of the form,
\begin{equation}
\label{2}
x'=\gamma(v) (x-vt)\,,
\end{equation}
with a coefficient $\gamma(v)$. 

\subsection{Case A: $\gamma(v)$ is an even function 
of $v$, i.e. $\gamma(v)=\gamma(-v)$}

From the perspective of $S'$, relative to which the frame $S$ moves backwards with velocity $-v$, 
the argument that led us to Eq.~(\ref{2}) now leads to
\begin{equation}
\label{3}
x=\gamma(v) (x'+vt')\,.
\end{equation}                             
Since $\gamma(v)=\gamma(-v)$, the $\gamma(v)$ appearing in Eq.~(\ref{3}) is the 
same as that appearing in Eq.~(\ref{2}).

Assume that the speed of light is equal to $c$ in both $S$ and $S'$, 
which is nothing but the postulate of constancy of the speed of light. 
In $S$, a light ray has trajectory $x = ct$, while, in $S'$, it has trajectory $x'=ct'$. 
Substituting these trajectories into Eqs.~(\ref{2}) and (\ref{3}) gives the following equations:
\begin{align}
\label{4a}
ct'&=\gamma(v) (c-v)t\,,\\
\label{4b}
ct&=\gamma(v) (c+v)t'\,,
\end{align}
which give
\begin{equation}
\label{5a}
\gamma^2(v)\big(1-v^2/c^2\big)=1\,.
\end{equation}
Eq.~(\ref{5a}) shows that $|v|<|c|$, which means that 
(under the assumptions of homogeneity of space-time, 
linearity of inertial transformation, and invariance of the speed of light) assuming $\gamma(v)$ 
is an even function of $v$ leads to the result that the relative speed between 
the two inertial frames is less than the speed of light.

Substituting the expression for $x'$ in Eq.~(\ref{2}) into Eq.~(\ref{3}) 
and using Eq.~(\ref{5a}), we get
\begin{equation}
\label{6}
t'=\gamma(v)\Big(t-\frac{v}{c^2}\, x\Big)\,,
\end{equation}
with the inverse transformation
\begin{equation}
\label{6b}
t=\gamma(v)\Big(t'+\frac{v}{c^2}\, x'\Big)\,.
\end{equation}
Eqs.~(\ref{4a}) and (\ref{4b}) show that $\gamma(v)$ is positive when $-c<v<c$, 
and from Eq.~(\ref{5a}), we obtain the usual Lorentz factor
\begin{align}
\label{7}
\gamma(v)=\frac{1}{\sqrt{1-v^2/c^2}}\,,\qquad-c<v<c\,.
\end{align}
Eqs.~(\ref{2}), (\ref{6}) and (\ref{7}) are the usual Lorentz transformation
connecting two subluminal frames.
Standard Lorentz transformation is characterized by its invariant action of the quadratic form:
$x'^2-c^2t'^2=x^2-c^2t^2$ and its Jacobian is: $J=1$. Thus, 
the standard Lorentz transformation is a proper transformation.

Assume a particle moves with constant velocity $u'$ in frame $S'$ which, in turn, moves with constant velocity $v$ ($|v|<c$) with respect to frame $S$. The velocity $u$ of the particle as seen in $S$ is just $u=x/t=\gamma(v)(x'+vt')/[\gamma(v)(t'+vx'/c^2)]$. Substituting  $x'=u't'$ into the expression for $u$ gives $u=(u'+v)/(1+u'v/c^2)$.

\subsection{Case B: $\gamma(v)$ is an odd function of $v$, i.e. $\gamma(v)=-\gamma(-v)$}

The same argument that led us to Eq.~(\ref{2}) now leads to 
\begin{equation}
\label{8}
x=-\gamma(v)(x'+vt')\,.
\end{equation}
Substituting the two trajectories $x = ct$ and $x'=ct'$ into Eqs.~(\ref{2}) and (\ref{8}), 
we obtain
\begin{align}
\label{9a}
ct'&=\gamma(v)(c-v)t\,,\\
\label{9b}
ct&=-\gamma(v)(c+v)t'\,,
\end{align}
which give
\begin{equation}
\label{10a}
\gamma^2(v)\big(v^2/c^2-1\big)=1\,.
\end{equation}
Eq.~(\ref{10a}) shows that $|v|>|c|$, indicating that (under the assumptions of homogeneity 
of space-time, linearity of inertial transformation, and invariance of the speed of light) 
assuming $\gamma(v)$ is an odd function of $v$ leads to the result that the relative speed 
between the two inertial frames is greater than the speed of light.

Substituting the expression for $x'$ in Eq.~(\ref{2}) into Eq.~(\ref{8}) 
and using Eq.~(\ref{10a}), we obtain
\begin{equation}
\label{11}
t'=\gamma(v)\Big(t-\frac{v}{c^2}\, x\Big)\,,
\end{equation}
with the inverse transformation
\begin{equation}
\label{11b}
t=-\gamma(v)\Big(t'+\frac{v}{c^2}\, x'\Big)\,.
\end{equation}
Eq.~(\ref{9a}) shows that $\gamma(v)$ is positive when $-\infty<v<-c$, and 
Eq.~(\ref{9b}) shows that $\gamma(v)$  is negative when $c<v<+\infty$, 
and combining with Eq.~(\ref{10a}), we obtain the generalized Lorentz factor for
superluminal motion
%%\begin{eqnarray}
\begin{align}
\label{12}
\gamma(v)=\left\{ 
\begin{array}{rl}
\displaystyle{\frac{1}{\sqrt{v^2/c^2-1}}}\,,\qquad &
\displaystyle{-\infty<v<-c}\\
\displaystyle{-\frac{1}{\sqrt{v^2/c^2-1}}}\,,\qquad &
\displaystyle{\quad\ c<v<+\infty}\ .\\
\end{array}
\right.
%%\end{eqnarray}
\end{align}
It is important to mention that this Lorentz factor is real-valued.
Eqs.~(\ref{2}), (\ref{11}) and (\ref{12}) are the generalized Lorentz transformations 
towards superluminal motion.

Assume a particle moves with constant velocity $u'$ in frame $S'$ which, in turn, 
moves with constant velocity $v$ ($|v|>c$) with respect to frame $S$. 
The velocity $u$ of the particle as seen in $S$ is just 
$u=x/t=-\gamma(v)(x'+vt')/[-\gamma(v)(t'+vx'/c^2)]$. 
Substituting $x'=u't'$ into the expression for $u$ gives $u=(u'+v)/(1+u'v/c^2)$, which is the same as in Case A.

Now some historical notes are in order to clarify the state of affairs.
From the historical point of view, it seems that~\citet{Synge} was the first who 
derived a real-valued Lorentz factor like the $\gamma$ in (\ref{12}a) for a point moving faster 
than the speed of light. 
\citet{Goldoni72,Goldoni73} derived such generalized Lorentz transformations 
with positive Lorentz factor~(\ref{12}a). 
\citet{Lord} and \citet{Sutherland} derived such generalized Lorentz transformations 
with negative Lorentz factor~(\ref{12}b). 
Many years later, \citet{HC} derived their two generalized Lorentz transformations 
which are the two transformations given earlier 
by~\citet{Goldoni72,Goldoni73,Lord} and \citet{Sutherland} 
and they agree with Eqs.~(\ref{2}), (\ref{11}) and (\ref{12}). 
It seems that \citet{HC} have overlooked the original generalized Lorentz transformations
given earlier by~\citet{Goldoni72,Goldoni73,Lord} and \citet{Sutherland}.
Thus, the ``Hill-Cox transformations'' are not at all new. 
Moreover,~\citet{HC13} continue to ignore the scientific literature in 
the re-publication of their results.
In addition, their statement that ``these independent derivations (\citet{HC} and \citet{Vieira}), obtained
from entirely distinct perspectives, mean that there is now some commonality of agreement in the basic
equations underlying superluminal motion" is of questionable veracity.

In addition, it is interesting to note that if $c$ plays the role of the speed of sound,
the Lorentz-like transformations with the Lorentz factor~(\ref{12}a) are in formal
agreement with the form of the Lorentz-like transformations for 
supersonic motion in aerodynamics (e.g. \citep{Miles,Shankara78}). 
As it seems \citet{Miles} was the first who derived a Lorentz-like transformation with 
a real valued Lorentz-factor for the transformation of 
the supersonic wing equations using the theory of aerodynamics.
%%%Once again, the story of Lorentz-like transformations is an old story.

The corresponding inverse transformations are given 
by Eqs.~(\ref{8}), (\ref{11b}) and (\ref{12}).
The transformations~(\ref{2}), (\ref{11}) and (\ref{12})
extend the Lorentz transformations towards
the regime $-\infty<v<-c$ and $c<v<+\infty$ in addition
to the ``classical'' regime $-c<v<c$.
Important enough to mention that for $c<v<+\infty$ as well as
for $-\infty<v<-c$,  Eq.~(\ref{12}) leads to real values
and not to imaginary values of the Lorentz factor $\gamma$ as often claimed 
in standard books on special relativity (e.g. \cite{Rindler80,Rindler91,LL,Inverno}).
Fig.~1 plots the curve for $\gamma$ as a function of $v/c$. 
The Lorentz factor $\gamma$ has thus been discussed in three regimes: 
the regime I shown in the present paper is $-\infty<v<-c$, 
the regime of special relativity, II, is $-c<v<c$, 
and the regime discussed by~\citet{HC}, III,  
lies in $c<v<+\infty$. 
Thus, \citet{Goldoni72,Goldoni73} derived extended Lorentz transformations
for the regime I, whereas \citet{Lord} and \citet{Sutherland} derived them for the
regime III.
It can be seen that $\gamma$ is singular at $v=-c$ and  at $v=c$. 
At $v=-c$, $\gamma$ possesses a singularity of $+\infty$, while $\gamma$ is discontinuous 
at $v=c$ since it jumps from $+\infty$ to $-\infty$. 
Hence, the region II, the domain of special relativity, 
is clearly separated from the other two regions.
In Fig.~1, it can be seen that $\gamma(v)$ is in the region II an even function
and in the regions I and III an odd function of $v$. 
It can be clearly seen in Fig.~1 that the region II is ``shielded'' from
regions I and III by two singularities at $v/c =-1$ and $v/c=+1$.
\begin{figure}[t]
\label{1}
\centerline{
\includegraphics[scale=0.8]{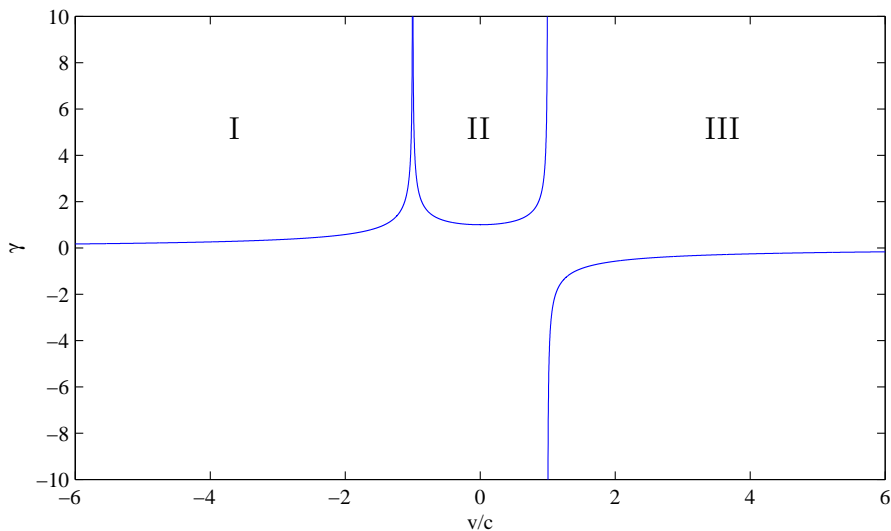}
\put(-250,150){I}
\put(-160,150){II}
\put(-70,150){III}
%%\put(-250,100){$\gamma$}
%%\put(-120,-4.0){$v/c$}
}
\caption{The curve for the Lorentz factor $\gamma$ as a function of $v/c$.}
\end{figure}

The limit of infinitely large velocity of
Eqs.~(\ref{11}) and (\ref{11b}) is now studied.
For the regime I the limit $v\longrightarrow-\infty$
in Eqs.~(\ref{11}) and (\ref{11b}) gives
\begin{align}
\label{lim1}
c\, t'=x\,,\qquad c\, t=x'\,
\end{align}
and for the regime III the limit $v\longrightarrow+\infty$,
Eqs.~(\ref{11}) and (\ref{11b}) are reduced to 
\begin{align}
\label{lim2}
c\, t'=x\,,\qquad c\, t=x'\,.
\end{align}
Also Eqs.~(\ref{2}) and (\ref{8}) give this limit.
Thus, the limit of infinitely large velocity gives 
in the regimes I and III
the exchange of space and time coordinates in $(1+1)$-dimensions. 
In addition, we want to mention that 
when $|v|>c$, the new coordinates $x'$ and $t'$ cannot longer be identified 
with a spatial and a temporal one, because they become timelike and spacelike, respectively. 
Thus, the new variables $x'$ and $t'$, although perfectly viable mathematical labels in 
spacetime, lack a proper chronological interpretation, which is what one wants from the coordinates in special relativity. 
This is particularly evident in the limit $|v|\longrightarrow\infty$, where 
equations~(\ref{lim1}) and (\ref{lim2}) imply that $t'$ is just a relabeling for $x$,
and $x'$ is just a relabeling for $t$.

In $(3+1)$-dimensions the Lorentz-like transformations
transform the wave operator
\begin{align}
\label{wave}
\pd_{xx}+\pd_{yy}+\pd_{zz}-\frac{1}{c^2}\, \pd_{tt}
\qquad
\text{to}
\qquad
-\pd_{x'x'}+\pd_{y'y'}+\pd_{z'z'}+\frac{1}{c^2}\, \pd_{t't'}
\end{align}
and the light-cone (equation of the wave front) 
\begin{align}
\label{LC}
x^2+y^2+z^2-c^2t^2 
=0
\qquad
\text{to}
\qquad
-x'^2+y'^2+z'^2+c^2 t'^2=0\,.
\end{align}
Thus, the wave operator and the light-cone are not invariant under the 
Lorentz-like transformations for superluminal motion.
The Lorentz-like transformations change the signature of the metric 
from $(+++-)$ to $(-+++)$.
This shows that the rotational symmetry in $(x,y,z)$-space (or isotropy of space) 
is lost for superluminal motion. This gives rise to a ``Mach cone'' in space-time
which divides space-time into two disconnected regions.
But this is not surprising since it is 
known from a Mach cone for supersonic speed in acoustics.
It can be seen that the surface to which Eq.~(\ref{LC}b) is asymptotic, namely
$ -x'^2+y'^2+z'^2=0$, is a Mach cone.

In $(1+1)$-dimensions the Lorentz-like transformations
transform the wave operator
\begin{align}
\label{wave-2}
\pd_{xx}-\frac{1}{c^2}\, \pd_{tt}
\qquad
\text{to}
\qquad
-\pd_{x'x'}+\frac{1}{c^2}\, \pd_{t't'}
\end{align}
and the light-cone 
\begin{align}
\label{LC-2}
x^2-c^2t^2=0
\qquad
\text{to}
\qquad
-x'^2+c^2 t'^2=0\,,
\end{align}
which looks like a formal exchange of space and time coordinates in $(1+1)$-dimensions. 
The signature of the metric is changed from $(+-)$ to $(-+)$ 
under the Lorentz-like transformations.
Accordingly, in the wave operator the $x$-axis becomes timelike and 
the $t$-axis spacelike.
The  Jacobian of the Lorentz-like transformations is: $J=-1$. Therefore, 
the Lorentz-like  transformations are improper transformations even in $(1+1)$-dimensions.
As a consequence the volume element is not invariant under Lorentz-like transformations:
$\de x'\de t' =J\, \de x\, \de t=-\de x\, \de t$.

\section{Conclusion and Discussion}

In the standard derivations of Lorentz transformations, the Lorentz factor $\gamma$ 
is assumed to be an even function of $v$ (see, e.g., \citep{Rindler80,Rindler91,2}). 
Following the definition in \citet{Inverno} and \citet{2}, 
this assumption is based on isotropy of space, 
i.e. space is non-directional, so that both orientations of the space axis 
are physically equivalent. 
The condition of isotropy of space is fulfilled only if $\gamma$ is 
an even function of $v$.
The case that $\gamma$ is an odd function of $v$
is not in agreement with the principle of relativity and with 
the condition of isotropy of space (see also~\citep{Vieira,Andreka}).
For special relativity, isotropy of space may be a sensible feature 
since both $v$ and $-v$ lie in the same 
regime II, but for the case of speeds greater than the speed of light, the velocity $-v$ 
lies in the entirely different regime I, and involves passing through the speed of light 
twice from the regime I to the regime III. 
In this note, it has been shown that, under the assumptions of homogeneity of space-time, 
linearity of inertial transformation, and invariance of the speed of light, 
the assumption of isotropy of space leads to the result that the relative speed between 
the two inertial frames is less than the speed of light, 
and the violation of isotropy of space results in a relative speed greater than the speed 
of light. 
From this connection between faster-than-light speeds and the issue of the violation of 
isotropy of space, the (real-valued) expressions for the Lorentz factor $\gamma$ 
in both the regimes I and III may be formally obtained. 
However, in the $(3+1)$-dimensional 
theory of special relativity Lorentz-like transformations 
(e.g. ``Hill-Cox transformations'') are not consistent because 
isotropy of space and rotational symmetry in space are broken.
In general, the Lorentz-like transformations 
(like ``Hill-Cox transformations'') are improper transformations due to $J=-1$. 
Thus, the wave operator, the light-cone and 
the volume element are not invariant under such Lorentz-like transformations.
We conclude that the ``Hill-Cox transformations'' are not new and they 
do not generalize Einstein's theory of special relativity in $(3+1)$-dimensions
as claimed by~\citet{HC,HC13}.

\section*{Acknowledgement}
M.L. gratefully acknowledges the grants obtained from the 
Deutsche Forschungsgemeinschaft (Grant Nos. La1974/2-1,  La1974/2-2, 
La1974/3-1). 
%%%The authors acknowledge the useful remarks and suggestions of the reviewers.

\end{document}